\documentclass[aps,pre,twocolumn,longbibliography,reprint]{revtex4-2}

\usepackage{graphicx}
\usepackage{amsmath}
\usepackage[colorlinks=true,hyperfootnotes=true,breaklinks=true]{hyperref}

\begin{document}

\title{Relationship between Heider links and Ising spins}

\author{Zdzis{\l}aw Burda}
\email{zdzislaw.burda@agh.edu.pl}
\thanks{ORCID~\href{https://orcid.org/0000-0002-9656-9570}{0000-0002-9656-9570}}

\author{Maciej Wo{\l}oszyn}
\email{woloszyn@agh.edu.pl}
\thanks{ORCID~\href{https://orcid.org/0000-0001-9896-1018}{0000-0001-9896-1018}}

\author{Krzysztof Malarz}
\email{malarz@agh.edu.pl}
\thanks{ORCID~\href{https://orcid.org/0000-0001-9980-0363}{0000-0001-9980-0363}}

\author{Krzysztof Ku{\l}akowski}
\email{kulakowski@fis.agh.edu.pl}
\thanks{ORCID~\href{https://orcid.org/0000-0003-1168-7883}{0000-0003-1168-7883}}

\affiliation{\href{https://ror.org/00bas1c41}{AGH University},
Faculty of Physics and Applied Computer Science,
al. Mickiewicza 30, 30-059 Krak\'ow, Poland}

\date{April 14, 2026}

\begin{abstract}
We show that the Heider model with an external field is equivalent, in the limit of structural balance, to the Ising model with nearest-neighbor interactions without an external field. More precisely, we claim that the signs of the Heider relations that maintain structural equilibrium in the system can be represented as nearest neighbor Ising spin products.
We demonstrate this explicitly for a complete graph and provide a general argument for an arbitrary graph.
A consequence of the equivalence is that the system of balanced Heider states undergoes a phase transition, inherited from the Ising model, at a critical value of the social field at which the fluctuations of edge magnetization are maximal.
\end{abstract}

\maketitle

\section{Introduction}

The Heider balance, or structural balance in signed graphs \cite{Heider,Cartwright_1956}, is a central problem in mathematical sociology \cite{Bonacich}.  Interest in this issue stems from the interpretation of positive and negative ties defined on graph edges as friendly and hostile relationships between entities (individuals) located at nodes. Research on Heiderian equilibrium and sign-based social networks is embedded in a relational sociology framework \cite{Donati_2011,Abbott2020}, in which the focus shifts from individuals to the relationships between them.

The problem of Heider balance is often discussed in connection with homophily, i.e., the tendency of a node to form a positive link with a node of similar characteristics \cite{McPherson_2001,Yap_2015,Lee_2025}. The outcome of this combination can be complex. Reference~\onlinecite{Deng_2016} demonstrates the competition between the homophily-driven formation of groups and the local adjustment towards balance. In Reference~\onlinecite{PhysRevLett.125.078302}, the system ability to achieve the balance depends on the number $G$ of attributes that determine the homophily process. It is demonstrated that above some critical value of $G$ depending on the size of the system $N$, the phase termed `paradise' (all positive links) is attained. A similar competition between homophily and balance has been considered in Reference~\onlinecite{Pham_2021}, with somewhat different results on the $G$-dependence of the final state; for large $G$, the balance is destroyed. We also note that an increase in social connectivity  has been found to produce fragmentation and polarization \cite{Pham_2020,Thurner_2025}. This discussion, apparently relevant for mathematical social sciences, seems far from a final conclusion.

Signed networks belong to a large family of complex systems, where a prescribed time evolution produces some stationary states.
In undirected networks, the evolution is directed as to minimize a work function, analogous to energy in physical systems. 
A physical system close to the network considered here is a spin glass \cite{Mezard_1987,RevModPhys.58.801,Mezard_2009}. 
In some models, signs of the exchange integrals of the magnetic interaction between nodes are randomly distributed. 
The key difference between our signed networks and a spin glass is that in the latter, magnetic moments $\pm 1$ of atoms at the network nodes undergo the evolution, while the links are fixed. 
On the contrary, in the signed network discussed here the dynamics applies to links, and spins are absent. Despite decades of efforts, physics of spin glasses remains a fertile field of research \cite{Ray_2026}. From this perspective, signed networks promise a path to understanding `a simpler complex system'.

On the other hand, sociological attempts to simulate complex networks are spread on a broader front of research \cite{Wasserman_Faust_1994}.
Still a tendency can be observed to focus on relations between individuals \cite{Selg_2020,Depelteau_2018}, rather than on individuals or groups as social entities. 
In computational models, the key effort is to identify a social process, and---if possible---a work function which could be minimized. 
This program can be executed only separately for a given social phenomenon. 
The present text is devoted to minimization of cognitive dissonance that drives the system to a balanced state. 

A balanced graph can be divided into two groups, with only positive links within each group and only negative links between the groups. This is the content of the structure theorem \cite{Cartwright_1956}. 
Consequently, in a balanced graph, each individual feels comfortable with a clear partition of the whole system into his or her enemies and friends; the related cognitive dissonance is removed \cite{Festinger_1957,Cooper_2019}. Whether the stationary state of a network is balanced depends on a particular system \cite{Szell_2010,Facchetti_2011,Doreian_2015,Liu_2020}. It is reasonable to view dissonance removal as one of the competing social processes \cite{Doreian_2009}.
 
The analysis of the equivalence of the Heider model and the Ising model \cite{Lenz1920,Ising1925,Brush_1967,Niss_2005} emphasizes the importance of spin models in the description of social systems. 
The Ising model is simple enough to be grasped and discussed in detail, both analytically and numerically. 
At the same time, it generates a reach spectrum of phenomena allowing to better understand:
models of opinion dynamics \cite{PhysRevE.94.062317,PhysRevE.103.L060301,Sousa2005,Malarz2003};
social segregation \cite{Stauffer_2007,Muller_2008};
crowd dynamics \cite{Korff2011,Krawczyk2014};
language dynamics \cite{Nettle_1999,Itoh_2004};
games theory \cite{Galam_2010,Leonidov_2024};
tax evasion \cite{Zaklan_2009,Giraldo-Barreto_2021};
modeling elections \cite{Devauchelle_2024,Korbel_2026};
large-scale social behaviors \cite{PhysRevX.9.011022,Wang_2025};
homophily in social relations \cite{PhysRevLett.125.078302,Li_2021};
psychological theories evaluation \cite{Maier_2024};
immigration, integration and ghetto formation \cite{Meyer-Ortmanns_2003};
the spread of infectious diseases \cite{Mello_2021,Benjamin_2023}
(see References~\onlinecite{Stauffer_2008,Mullick_2025} for comprehensive reviews), to mention only several in the field of sociophysics \cite{Galam_2012,Stauffer_2013,Sen_2014,Schweitzer_Sociophysics,da_Luz_2023}.

The concept of the applied field is useful in social psychology \cite{Lewin_1947,Helbing_1994}, as it can represent incentives to improve or deteriorate interpersonal relationships. An external field in this role has been integrated into the Heider model in References~\onlinecite{Oloomi_2023,2512.00567}. The field and thermal noise complement the toolbox of statistical mechanics \cite{Stanley}.  
Our aim here is to demonstrate the equivalence of the balanced system in the presence of an external field and the standard Ising system without a field.

\section{Methods}

Consider a social network represented as a complete graph on $N$ nodes, equipped with spin variables $s_{ab}=\pm 1$ for all pairs of nodes, reflecting the type of relationship between $a$ and $b$, which is either friendly, $s_{ab}=+1$, or unfriendly, $s_{ab}=-1$. An elementary cycle $abc$ is structurally balanced if the product of the edge signs is $s_{ab}s_{bc}s_{ca}=+1$. If all elementary cycles are balanced, then the network is balanced.
The network has $N_e=\binom{N}{2}$ edges and $N_t=\binom{N}{3}$ triangles that form elementary cycles. 

A balanced network has the following topological property \cite{Cartwright_1956}: the parity $\pi$ of any path $\gamma=(a_1a_2, a_2a_3,\ldots, a_na_{n+1})$, defined as the product of the signs of the edges that form the path
\begin{equation}
  \pi(\gamma) = \prod_{i=1}^{n} a_ia_{i+1}
\end{equation}
depends only on the end-points. In other words, if  two paths $\gamma_{ab}$ and $\gamma'_{ab}$ have the same end-points $a$ and $b$, then $\pi(\gamma_{ab})=\pi(\gamma'_{ab})$ holds independently of the shape and length of the paths. In particular, the parity of any cycle $\gamma_c$ is $\pi(\gamma_c)=+1$. 

An interesting question is how the properties of structural equilibrium depend on the relative frequency of friendly and unfriendly relationships in the network.
To investigate this problem, let us introduce a field $h$ that breaks the symmetry between friendly and unfriendly relations. 
We use a sign convention according to which a positive $h$ favors friendly relations, and a negative $h$ favors unfriendly relations. The symmetry is restored for $h=0$. 
The system can be described by the partition function  
\begin{equation} \label{eq:Z}
Z = \sum_{\{s_{ab}=\pm 1\}} \exp(-\beta U),
\end{equation}
where $s_{ab}=s_{ba}$ are spin-like variables $s_{ab}=\pm 1$,
and the energy is
\begin{equation} \label{eq:U}
U = -\varepsilon P - h M = -
\frac{\varepsilon}{n_t}  \sum_{(abc)} s_{ab}s_{bc}s_{ca}
- \frac{h}{n_e} \sum_{(ab)} s_{ab},
\end{equation}
where $n_t = N_t/N = (N-1)(N-2)/6$ is the density of triangles per node, and $n_e=N_e/N = (N-1)/2$ is the density of edges per node. 
The first sum runs over all triangles, and the second one runs over the links of the network. 
In physical terms, $M$ can be called the edge magnetization, and $m=M/N$ the magnetization density. 
Positive/negative $M$ indicates friendly/unfriendly relations. The term $P$ sums the parities of the elementary cycles; $p=P/N$ is the average parity per cycle. There are two coupling
constants $\varepsilon$, $h$ in Equation~\eqref{eq:U}. The first is responsible for the average parity of elementary cycles, and the second
is responsible for the relative prevalence of friendly and unfriendly relationships. The normalization factors $1/n_t$ and $1/n_e$ in front of the sums \eqref{eq:U} are chosen so that both terms $P$ and $M$ are extensive in $N$.
The partition function \eqref{eq:Z} describes a thermodynamic system with temperature $T=1/\beta$. The temperature sets the 
scale for the coupling constants $\varepsilon$ and $h$.
The partition function 
\begin{equation} \label{eq:Z'}
Z = \sum_{\{s_{ab}=\pm 1\}} 
\exp \left(\frac{\varepsilon'}{n_t}  \sum_{(abc)} s_{ab}s_{bc}s_{ca} + \frac{h'}{n_e} \sum_{(ab)} s_{ab}\right)
\end{equation}
depends on $\varepsilon, h$ and $T$ via effective parameters $\varepsilon' = \varepsilon/T$, $h' = h/T$.
In the limit $\varepsilon' \rightarrow \infty$,
the system is structurally balanced.
When the sum $\{s_{ab}=\pm 1\}$ in Equation~\eqref{eq:Z'} is restricted to the subspace of balanced states
$\{s_{ab}=\pm 1\}_{B}$, such that $s_{ab}s_{bc}s_{ca}=1$
for all triangles, the partition function reduces to
\begin{equation} \label{eq:ZB}
Z = e^{\varepsilon' N} \sum_{\{s_{ab}= \pm 1\}_B} 
\exp\left(\frac{h'}{n_e} \sum_{(ab)} s_{ab}\right).
\end{equation}
The constraints $s_{ab}s_{bc}s_{ca}=1$ can be resolved
by choosing all edge spins $s_{ab}$ in the following form  
\begin{equation} \label{eq:eq}
    s_{ab} = \sigma_a \sigma_b,
\end{equation} 
where $\sigma_a=\pm 1$ are spin-like variables associated with all the nodes of the network. The partition function \eqref{eq:ZB} can thus be written as
\begin{equation} \label{eq:ZI}
Z = e^{\varepsilon' N} \sum_{\{\sigma_{a}= \pm 1\}} \exp\left( \frac{h'}{n_e} \sum_{(ab)} \sigma_a \sigma_b\right) = e^{\varepsilon' N} Z_{I}.
\end{equation}
We see that the partition function for structurally balanced
states in the presence of the external field is equal, up to the factor $e^{\varepsilon' N}$, to the partition function $Z_I$ of the nearest-neighbor Ising model with no external field. 
The field that breaks the symmetry between the probabilities of friendly and unfriendly relations in the Heider model plays the role of the spin coupling constant in the Ising model. 
The factor $e^{\varepsilon' N}$ is independent of $h'$, so $\partial_{h'} \ln Z= \partial_{h'} \ln Z_{I}$. The derivative $\partial_{h'} \ln Z =\langle M \rangle$ is equal to the average edge magnetization in the Heider model, while $\partial_{h'} \ln Z_I =\langle E \rangle$ is equal to the average  energy in the Ising model,
where $E=\sum_{(ab)} \sigma_a \sigma_b$. 
The second derivative is equal to the variance of magnetization $\partial^2_{h'h'} \ln Z = \langle M^2 \rangle - \langle M\rangle^2$, which, in turn, is equal to the variance of energy $\partial^2_{h'h'} \ln Z_I = \langle E^2 \rangle - \langle E\rangle^2$ in the Ising model. 

\begin{figure}
\includegraphics[width=\columnwidth]{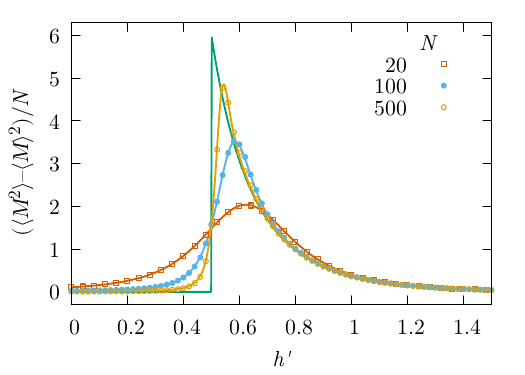}
\caption{\label{fig:phi_hh} Susceptibility (normalized variance)
$\chi_m=\left(\langle M^2 \rangle - \langle M \rangle^2\right)/N$ for balanced states in the Heider model on the complete graph $K_N$. The solid lines were calculated analytically for $N=20,100,500$ (red, blue, yellow), thanks to the equivalence with the Ising model, from \eqref{eq:Zfs1} for finite $N$. 
The green solid line was calculated from the saddle point equations (\ref{eq:sp1}, \ref{eq:sp2}) for $N=\infty$. 
The symbols, with the corresponding colors, were obtained using Monte Carlo simulations of the Heider model with a very large $\varepsilon'=1000$ to mimic $\varepsilon'=\infty$, where the system
is in Heider balance, and the equivalence to the Ising model is exact}
\end{figure}

The Ising model on a complete graph can be solved analytically, thanks to the relation $E=(S^2-N)/2$
between the energy $E=\sum_{(ab)} \sigma_a \sigma_b$ and the magnetization $S=\sum_a \sigma_a$, which holds on this graph. One can easily find that the partition function $Z_I$  \eqref{eq:ZI} is 
\begin{equation} \label{eq:Zfs1}
    Z_I = \sum_{k=0}^{N} \binom{N}{k} \exp \left[ \frac{h'}{N-1}\left((2k-N)^2-N\right)\right] .
\end{equation}
The summation index $k$ can be interpreted
as the number of positive spins, and the binomial symbol
represents the number of configurations that have $k$ positive spins.
The coefficient $N-1$ is obtained by substituting $n_e=(N-1)/2$. The sum \eqref{eq:Zfs1} also has an integral representation
\begin{equation} \label{eq:Zfs2}
\begin{split}
   & Z_I =  \sqrt{\frac{(N-1)h'}{4\pi}} \exp\left
    (-\frac{Nh'}{N-1}\right) \times \\
   &  \int_{-\infty}^{+\infty} dx \exp\left[\frac{1}{4}h'x^2+ 
      N\left(-\frac{1}{4}h' x^2 + \ln\left(2\cosh(h'x)\right)\right) \right]
\end{split}
\end{equation}
which can be obtained by the Hubbard–Stratonovich transformation applied to $\exp \left(h'/n_e (\sum_a \sigma_a)^2\right)$ in Equation~\eqref{eq:ZI}. The integral representation is particularly useful in the thermodynamic limit $N\rightarrow \infty$ to determine the asymptotic
behavior of the partition function, which, in the leading
order, grows exponentially with the coefficient 
\begin{equation}
    \phi_I = \lim_{N\rightarrow \infty } \frac{\ln Z_I}{N}.
\end{equation}
The steepest descent method gives
\begin{equation} \label{eq:sp1}
    \phi_I = -\frac{1}{4}h' x^2_* + \ln\left(2\cosh(h'x_*)\right),
\end{equation}
where $x_*=x_*(h')$ is a solution to the saddle point equation
\begin{equation} \label{eq:sp2}
    x_* = 2 \tanh(h' x_*).
\end{equation}
For $h'\le 1/2$, the last equation gives $x_*=0$, while for $h'>1/2$, it yields positive $x_*$, which for small positive $\Delta h = h'-1/2$ behaves as  $x_*= \sqrt{24} \sqrt{\Delta h} + o(\sqrt{\Delta h})$.
The second derivative $\partial^2_{h'h'} \phi_I$ is discontinuous for $h'=1/2$, where it jumps from $0$ to $6$. The result is shown in Figure~\ref{fig:phi_hh}.
The limit curve for $N\rightarrow \infty$ is  obtained using Equations~\eqref{eq:sp1} and \eqref{eq:sp2} while the finite size results are obtained from Equation~\eqref{eq:Zfs1}.

\section{Results}

We can now compare these results with the results of the Monte Carlo simulations of the Heider model. 
We will use the heat bath algorithm with two types of updates: link updates and vertex updates. 
To perform a link update, we randomly choose a link $ab$, compute the local field derived from all `staples' (complements of elementary cycles) containing the link $\xi_{ab} = \frac{h'}{n_e}+\frac{\varepsilon'}{n_t} \sum_{c\ne a,b} s_{ac}s_{cb}$, 
and choose a new spin edge value $s_{ab}$ with probability 
$e^{s_{ab} \xi_{ab}}/(e^{\xi_{ab}}+e^{-\xi_{ab}})$. 
To perform a vertex update, we randomly choose a vertex $a$, compute a magnetic local field $\zeta_a = \frac{h'}{n_t} \sum_b s_{ab}$ coming from all edges emerging from $a$, choose $\sigma_a = \pm 1$ with probability $e^{\sigma_a \zeta_a}/(e^{\zeta_a}+e^{-\zeta_a})$, and set $s_{ab} \rightarrow \sigma_a s_{ab}$ for all edges attached to 
$a$. Vertex updates do not change the contribution from elementary cycles, since in each cycle, an even number of edges changes the sign. Vertex updates improve the algorithm's efficiency for large $\varepsilon'$, where the link updates have a low acceptance.

\section{Discussion}

A comparison of the Monte Carlo results with the theoretical predictions resulting from the equivalence of the Heider model
for $\varepsilon' \rightarrow \infty$ with the Ising model is presented in Figure~\ref{fig:phi_hh}. 
Both models give exactly the same results. One can also see that the normalized variance $\chi_m$ is a function of the field $h'$, 
which is a good scaling variable in the limit $\varepsilon'\rightarrow \infty$. On the other hand, for $\varepsilon'=0$, the normalized variance can easily be found to be $\chi_m = \cosh^{-2}(h'/n_e)/n_e^2$. This follows from the independence of edge signs for $\varepsilon'=0$. So we see that in this case $h'/n_e$ rather than $h'$ is a good scaling variable. The corresponding
curve $\chi_m$ plotted as a function of $h'$ will broaden as $N$ increases.

\begin{figure}
\includegraphics[width=\columnwidth]{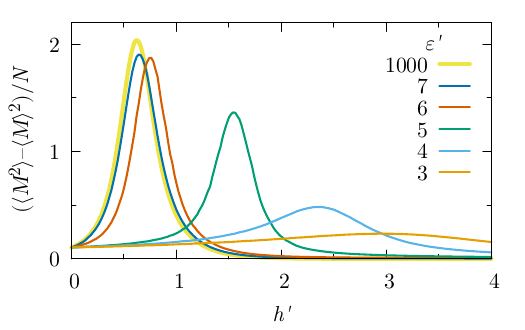}
\caption{\label{fig:N20}
Evolution of the normalized variance
$\chi_m$ with $\varepsilon'$, for $N=20$. The curves are obtained
by Monte Carlo simulations. For $N=20$, the critical value is $\varepsilon'_{cr} \approx 6$, see Figure~\ref{fig:epscr}. 
For $\varepsilon' > \varepsilon'_{cr}$, the curves quickly converge to a
limit curve. For $\varepsilon' < \varepsilon'_{cr}$, the curves broaden when $\varepsilon'$ increases; they also broaden when $N$ increases 
although this is not shown in the figure}
\end{figure}

The situation is illustrated in Figure~\ref{fig:N20} where we plot $\chi_m$ vs. $h'$ for different values of $\varepsilon'$ for
$N=20$. 
The curves for $\varepsilon' \ge 6$ have comparable widths and a more or less constant position. 
They only slightly deviate from the limiting curve, while the curves for $\varepsilon' \le 3$ are much wider and have a moving mode position. 
A similar picture is observed for other sizes of the system $N$. 

\begin{figure}
\includegraphics[width=\columnwidth]{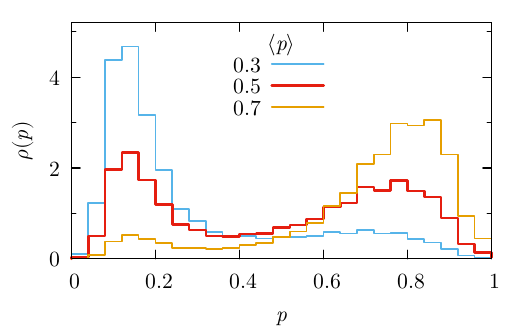}
\caption{\label{fig:hist}Histogram of the probability density function $\rho(p)$ with the bin size $0.04$ for $N=20$, for three different values of $\varepsilon'=5.88$, $6.02$, and $6.18$. The mean values (which correspond to the positions of the histogram mass centers) are $\langle p \rangle = 0.3$, $0.5$, and $0.7$ respectively. The functions have a bimodal shape typical for the first order phase transition.
Changing $\varepsilon'$ changes proportions of the two phases.  
The pseudo-critical value $\varepsilon'_{cr}$ is defined as a value of $\varepsilon'$ for which $\langle p \rangle=1/2$}
\end{figure}

The picture is correlated with the phase diagram of the model. 
The model undergoes a first order phase transition at a critical value of $\varepsilon'$, where the average parity of elementary loops $p$ has a discontinuity when plotted as a function of $\varepsilon'$. 
In the case of finite $N$, instead of a discontinuity, a sharp increase is observed with a slope of $1/N$ in a narrow interval.
At the transition point, the histogram of $p$ has a bimodal structure consisting of two peaks corresponding to coexisting phases.
The appearance of a bimodal histogram, presented in Figure~\ref{fig:hist}, is a strong indication of a first-order phase transition. 
When the value $\varepsilon'$ changes, so do the proportions of the peaks. 
There are different ways to define the pseudo-critical value \cite[p.~162]{Guide_to_Monte_Carlo_Simulations_2009}. 
We find it convenient to define it as the value of $\varepsilon'$ at which the mean of the distribution of $p$ is equal to $1/2$, which only occurs when the two phases coexist (in comparable proportions).

\begin{figure}
\includegraphics[width=\columnwidth]{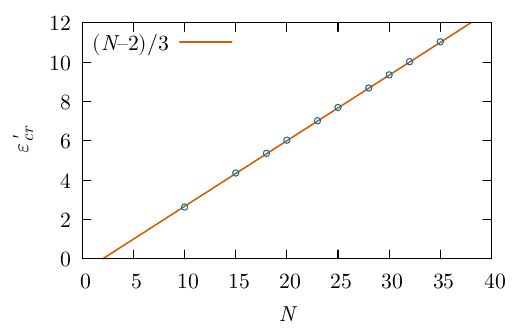}
\caption{\label{fig:epscr} The pseudo-critical value 
$\varepsilon'_{cr}$ for different values of $N$, for $h'=0$. 
The phase transition is first order. 
At the phase transition two phases co-exist: one with $p<1/2$
and the other with $p>1/2$, see Figure~\ref{fig:hist}. The pseudo-critical
values $\varepsilon'_{cr}$ shown as points for different $N$, are obtained
from the condition that the mean (the center of mass of the histogram)
is $\langle p \rangle =1/2$. The 
line $\varepsilon'_{cr}=(N-2)/3$, drawn as a solid line, 
passes through the points}
\end{figure}

Using this definition, we found numerically that the pseudo-critical value grows linearly with $N$ roughly as: $\varepsilon'_{ cr}\approx (N-2)/3$; 
see Figure~\ref{fig:epscr}. The estimated dependence on $N$ does not differ significantly from the mean field result, which also predicts that
for large $N$ the pseudo-critical value increases linearly with $N$ as
$\varepsilon'\approx 0.286 (N-1)$ for $h'=0$ \cite{1911.13048}. 

The parameter $h'$ is a good scaling variable in a part of the ordered phase where $p\approx 1$, that is, for $\varepsilon' \gg \varepsilon'_{cr} \sim O(N)$. This is also part of the phase diagram where the equivalence with the Ising model holds. 
On the other hand, the combination $h'/n_e$ is a good scaling variable in a part of the disordered phase where $p\approx 0$, that is, for $\varepsilon \ll \varepsilon'_{cr} \sim O(N)$; for example, for any finite $\varepsilon'$.

To summarize, we have shown that the Heider model on the complete graph is equivalent to the Ising model in the limit $\varepsilon' \to \infty$. 
In this limit, the system is in Heider balance, leaving limited freedom for the link signs.
Generally, there are $2^{N_e}$ states, but only $2^{N}$ balanced states. As a consequence, the entropy, being generally proportional to $N_e$, shrinks to become  proportional to $N$ in the limit $\varepsilon' \to \infty$. 
The shrinkage factor $N/N_e=1/n_e$ reappears in the scaling variable, which is $h'/n_e$ for finite $\varepsilon'$, and changes to $h'$ in the limit $\varepsilon' \to \infty$. 

Although equivalence between Heider and Ising models may seem trivial, its consequences are not:
\begin{itemize}
\item First of all, for $h = 0$, the model has local symmetry. This type of symmetry is highly non-trivial and in the language of field theory it is called gauge symmetry. Using this equivalence, one can compute the volume of the symmetry group (gauge orbit).
\item Second, for $h \ne 0$, local symmetry is broken, but the equivalence allows us to show that the partition function of the Heider model can be mapped onto the partition function of the Ising model, in which the social field $h$ corresponds to the reciprocal of the temperature, and the susceptibility in the Heider model corresponds to the specific heat in the Ising model due to the social field. To the best of our knowledge, the equivalence of these two models was not known.
\end{itemize}

The equivalence of the balanced Heider states with the Ising states is a direct consequence of Equation~\eqref{eq:eq}, which resolves simultaneously the constraints $\pi(\gamma_e)=1$ for all elementary cycles. 
This equivalence holds for the Heider model on any graph, not only on the complete graph. 
The phase transition in the Heider model, for balanced states, is inherited from the phase transition of the Ising model. 
The order of this transition thus depends on the dimensionality of the problem. In particular, the susceptibility $\chi_m=\partial^2_{h'h'} \ln Z$ exhibits a logarithmic divergence at the critical value of the field in two dimensions \cite{2512.00567}, and a power-law divergence $\chi_m \sim |\Delta h|^{-\alpha}$, with $\alpha = 0.11008708(35)$, inherited from the three dimensional Ising model \cite{Chang_2025}.

The equivalence of field-dependent balanced systems with the Ising model is a new example of the application of statistical mechanics \cite{Stanley} to relational sociology \cite{Donati_2011}. It extends the list of problems in which the language and tools of statistical mechanics have proven to be very useful in describing social realms quantitatively
\cite{PhysRevLett.130.037401,PhysRevLett.132.077401}. 

Last but not least, it would be interesting to further explore the relationship between the models in order to find an interpretation of the critical value of the field in terms of social relations. 
Moreover, discrete states of interpersonal relations, obtained as fixed points in nonlinear dynamics \cite{PhysRevLett.133.127402,kk}, can be supplemented with ubiquitous noise to provide more general models of social phenomena.

\begin{acknowledgments}
This research was supported by a subsidy from the Polish Ministry of Science and Higher Education.
\end{acknowledgments}


%
\end{document}